\documentclass[prd,aps,preprintnumbers,showpacs,twocolumn,superscriptaddress,floatfix,nofootinbib]{revtex4-1}

\usepackage{latexsym} 
\usepackage{amssymb} 
\usepackage{amsfonts}
\usepackage{amsmath} 
\usepackage{bm} 
\usepackage{color}
\usepackage{times} 
\usepackage{units}
\usepackage{hyperref}

\usepackage[utf8x]{inputenc}
\usepackage{graphicx} 
\usepackage[squaren]{SIunits} 

\newcommand{\Eref}[1]{Eq.~\eqref{#1}} 
\newcommand{\Fref}[1]{Fig.~\ref{#1}} 
\newcommand{\Sref}[1]{Sec.~\ref{#1}} 
\newcommand{\Tref}[1]{Table~\ref{#1}} 

\newcommand{\Mbar}{\ensuremath{M_\text{B}}}

\newcommand{\Mbh}{\ensuremath{M_\text{BH}}}
\newcommand{\Mdisk}{\ensuremath{M_\text{D}}}
\newcommand{\Minf}{\ensuremath{M_\infty}}
\newcommand{\Jadm}{\ensuremath{J_\text{ADM}}}
\newcommand{\Jbh}{\ensuremath{J_\text{BH}}}

\newcommand{\NSspin}{\ensuremath{\chi_{_\text{NS}}}} 
\newcommand{\BHspin}{\ensuremath{\chi_{_\text{BH}}}} 
 
\newcommand{\rcirc}{r_c}

\newcommand{\Msol}{\ensuremath{M_\odot}}

\newcommand{\msec}{\usk\milli\second} 
\newcommand{\parsec}{\mathrm{pc}}

\begin{document}


\title{Properties of hypermassive neutron stars formed in mergers of
spinning binaries}

\author{Wolfgang \surname{Kastaun}} 
\affiliation{Physics Department, University of Trento, 
via Sommarive 14, I-38123 Trento, Italy}
\affiliation{ 
Max-Planck-Institut für Gravitationsphysik, 
Albert Einstein Institut, Am Mühlenberg 1, 
14476 Potsdam, Germany}

\author{Filippo \surname{Galeazzi}}
\affiliation{ 
Institut für Theoretische Physik, Max-von-Laue-Straße 1, 
60438 Frankfurt, Germany}


\begin{abstract}
We present numerical simulations of binary neutron star mergers, 
comparing irrotational binaries 
to binaries of NSs rotating aligned to the orbital angular momentum.
For the first time, we study spinning BNSs employing nuclear physics 
equations of state, namely the ones of
Lattimer and Swesty as well as Shen, Horowitz, and Teige.
We study mainly equal mass systems leading to a hypermassive neutron 
star (HMNS), and analyze in detail its structure and dynamics. 
In order to exclude gauge 
artifacts, we introduce a novel coordinate 
system used for post-processing.
The results for our equal mass models show that the strong radial 
oscillations of the HMNS modulate the instantaneous frequency of 
the gravitational wave (GW) signal to an extend that leads
to separate peaks in the corresponding Fourier spectrum.
In particular, the high frequency peaks which are often attributed 
to combination frequencies 
can also be caused by the modulation of the $m=2$ mode frequency in the merger 
phase. As a consequence for GW data analysis, 
the offset of the high frequency peak does not 
necessarily carry information about the radial oscillation frequency.
Further, the low frequency peak in our simulations is dominated by 
the contribution of the plunge and the first 1--2 bounces.
The amplitude of the radial oscillations depends on the initial
NS spin, which therefore has a complicated influence on the spectrum.
Another important result is that HMNSs can consist of a slowly 
rotating core with an extended, massive envelope rotating close to 
Keplerian velocity, contrary to the common notion that a rapidly rotating 
core is necessary to prevent a prompt collapse.
Finally, our estimates on the amount of unbound matter 
show a dependency on the initial NS spin, explained by
the influence of the latter on the amplitude of radial oscillations, 
which in turn cause shock waves. 
\end{abstract}
  
\maketitle

\section{Introduction}
\label{sec:intro}

Relativistic numerical simulations of binary neutron star (BNS) mergers are
bringing the fate of the remnant object into increased focus (see 
e.g.~\cite{Baiotti08, Stergioulas2011b, Sekiguchi2011, Hotokezaka2011,
Paschalidis2012, Bauswein:2010dn, Bauswein2013, Bauswein2013b, 
Hotokezaka2013, Hotokezaka2013c, Takami2014} and references therein). 
Such simulations have revealed that hyper- (or supra-) massive 
neutron stars (HMNS; see~\cite{Baumgarte00bb} for precise definitions) are
the most likely outcomes of BNS mergers with equations of state (EOS)
that allow a maximum mass for non-rotating neutron stars (NS) in the 
range $2.6$--$2.8\usk M_{\odot}$. 

Among the plethora of interesting physics that can be extracted from 
the numerical analysis, the understanding of the features encoded in 
the gravitational radiation produced by the complex 
dynamics of the HMNS is perhaps the chief motivation driving most 
simulations. The prospects for detecting such signals with 
ground-based laser interferometers are continuously improving, 
see~\cite{Harry2010, Accadia2011_etal, Aso:2013}. The study of the 
gravitational radiation frequency spectra of HMNS might in particular open 
the possibility of inferring the masses and radii of neutron stars 
through common techniques of asteroseismology~\cite{Bauswein2011, 
Hotokezaka2011, Stergioulas2011b, Takami2014, Bauswein2014}, a 
long-standing issue in relativistic astrophysics. 
The extraction of two mode frequencies through a successful gravitational
wave (GW) detection might help to further tighten existing constraints on the 
EOS for high-density matter, such as the recent observations
~\cite{Demorest2010,Antoniadis2013} of $\sim 2~M_{\odot}$
neutron stars.

Important steps in this direction were taken 
by~\cite{Stergioulas2011b} who studied the oscillation modes of the 
HMNS as an isolated gravitating fluid. It was found that the fluid 
oscillations correspond directly to peaks in the GW 
spectrum, in which the most salient feature is the quadrupole ($m=2$) 
oscillation mode which appears as a triplet, with side bands caused 
by the non-linear coupling to the fundamental quasi-radial ($m=0$) 
mode. However, this interpretation implies oscillations with 
essentially fixed frequencies. During the plunge and merger phase 
however, the system is still highly non-linear. As already pointed 
out in \cite{Hotokezaka2013}, this often results in strong modulations
of the mode frequencies themselves. We will show that the high 
frequency peak does not necessarily correspond to a combination 
frequency, but can also be caused directly by the $m=2$ perturbations 
during the merger phase, where the frequency shortly reaches higher 
values. 

A more recent study~\cite{Takami2014} has focused 
on the low-frequency peak of the spectrum, showing that it satisfies 
an essentially universal relation with the compactness of the stars in 
the binary so that a robust method to constrain the EOS can be derived 
by combining the information from the two peaks. The low frequency peak 
was interpreted in~\cite{Takami2014} as the result of a double core 
structure during the merger. We support that view by proving that the 
low frequency peak in our simulations is due to the GW signal during 
the plunge and the first bounces.

In contrast to binary black hole (BH) mergers, only few simulations investigate the 
influence of the initial NS spin \cite{Tsatsin2013, Bernuzzi2013prd, 
Kastaun2013}. The reason is that consistent initial data for spinning 
NS binaries has become available only recently \cite{Tichy11}. In our 
previous work \cite{Kastaun2013}, we successfully used approximated 
initial data for spinning binaries, which we also employ here. In 
contrast to existing studies which assume simplified analytic EOSs, we 
evolve spinning binaries with nuclear physics EOSs. In 
\cite{Kastaun2013}, we studied the influence of the NS spin on the 
spin of the BH. In this work, we also provide some
data points for models with nuclear physics EOSs. Our main focus is 
however on the cases where a HMNS is formed. In \cite{Bernuzzi2013prd}, 
a dependency of the $m=2$ mode frequency on the initial NS spin was 
discovered. We investigate the impact on the GW spectrum in more 
detail, revealing a complicated picture which makes it difficult to 
deduce the spin from the GW signal.

We also investigate the influence of the initial NS spin on the 
ejected matter for the first time. The amount, composition, and 
temperature of matter ejected in BNS mergers is of considerable 
astrophysical interest since $r$-process nucleosynthesis in the ejected 
matter is one explanation for the abundance of heavy elements in 
the universe. Our study does not include neutrino radiation and hence 
cannot provide realistic values for the composition of ejected matter. 
However, our estimates indicate a significant impact of the spin on the 
amount of ejected matter, mediated by the influence  on the radial 
oscillation amplitude. 

A very important quantity in BNS merger simulations is the threshold mass 
above which the merger results in a prompt collapse to a BH. In particular, 
current models linking BNS mergers to short gamma ray bursts require the 
formation of a BH. In \cite{Bauswein2013}, the threshold mass was 
computed for a wide range of EOSs, using the conformal flatness 
approximation. We confirm those results in full GR for the two EOS considered 
in our work.

A related question is what exactly prevents the immediate collapse of 
HMNSs to a BH. The exact contributions of differential rotation and/or 
thermal effects are still largely unconstrained. It is known that 
isolated NSs can support large masses if they follow a $j$-const 
rotation law for which the core rotates more rapidly 
\cite{Baumgarte00bb}. Such configurations are the standard model 
for HMNSs formed in mergers. Some results presented in 
\cite{Shibata05c} on the other hand indicate a slower central 
rotation rate. In this paper we study the rotation profiles in detail and 
show that HMNSs can indeed exhibit a rotation profile where the core 
rotates slowly, while the outer layers rotate only slightly below Kepler 
velocity. The question whether the core or the outer layers stabilize 
a HMNS is important for estimating the lifetime. For example, it was 
shown in \cite{Paschalidis2012} that an artificial cooling mechanism 
acting on the whole system can reduce the lifetime of HMNSs (with a 
simplified analytic EOS). Neutrino cooling on the other hand will 
probably have a stronger effect on the outer, optically thin layers 
of a HMNS. The rotation profile is also important for models describing 
the amplification of the magnetic field (which is not included in our 
work), such as \cite{Siegel2013, Kiuchi2014}.

The paper is organized as follows: \Sref{sec:numerics} describes technical 
details of our simulations such as the computational framework and a 
novel coordinate system used for post-processing. Our binary NS models
are described in \Sref{sec:models}. The results of our simulations 
are presented in \Sref{sec:results}. In particular, \Sref{sec:inspiral} 
summarizes the inspiral phase and the outcome of the merger, \Sref{sec:hmns} 
addresses the structure and evolution of the HMNSs, \Sref{sec:gw} relates the 
GW signals and spectra to the dynamics of the HMNS, and \Sref{sec:ejecta} 
is dedicated to matter ejection. Finally, a summary and a discussion is 
provided in \Sref{sec:summary}.

\section{Numerical Methods}
\label{sec:numerics}

\subsection{Evolution framework}
\label{sec:evolcode}
All simulations in this work rely on the \texttt{WhiskyThermal} 
code described in \cite{Galeazzi2013, Alic2013}
to evolve the general relativistic
hydrodynamics equations (see also~\cite{Font08}).
The code implements finite-volume, high-resolution shock-capturing 
methods, and makes use of nuclear physics EOSs in tabulated form,
including thermal and composition effects.
We do not include neutrino radiation in our study, and simply 
advect the electron fraction together with the fluid.

The spacetime is evolved by the \texttt{McLachlan} 
code~\cite{Brown2007b}, which is part of the publicly available 
Einstein Toolkit~\cite{loeffler_2011_et}. All codes are based on the \texttt{Cactus} 
computational framework.
The \texttt{McLachlan} code employs high order finite-difference 
methods and offers different options for the formulation of the 
Einstein equations. Instead of the popular BSSNOK 
formulation~\cite{Nakamura87, Shibata95, Baumgarte99}, 
we chose the CCZ4 formulation described in 
\cite{Alic:2011a, Alic2013} because of its constraint-damping 
abilities. 

As gauge conditions, we apply the ``$1+\log$''-slicing 
condition~\cite{Bona94b} for the lapse function together 
with the hyperbolic $\tilde{\Gamma}$-driver 
condition~\cite{Alcubierre02a} for the shift vector.
At the outer edge of the 
computational domain, we employ Sommerfeld radiative boundary 
conditions for the spacetime. 
Furthermore, we enforce reflection symmetry with respect to the 
orbital plane. We do however not enforce $\pi$-symmetry,
to avoid suppressing potential physical instabilities.

Our simulations make use of moving box mesh refinement, 
provided by the \texttt{Carpet} code~\cite{Schnetter-etal-03b}.
In detail, we use six refinement levels during the inspiral,
with the two finest ones tracking the movement of the NSs.
Close to the merger, when the moving refinement regions are 
overlapping, they are replaced with fixed refinement regions
fully containing the previous ones. When a collapse to a BH is 
imminent, we activate another refinement level in order to 
improve the accuracy of the BH properties.
To ascertain numerical accuracy of the evolution framework, 
we rely on the tests already performed in \cite{Galeazzi2013, Alic2013},
which include an expensive convergence test of a binary merger 
simulation.

We detect the formation of apparent horizons utilizing the module 
\texttt{AHFinderDirect}~\cite{Thornburg2003:AH-finding} from
the Einstein toolkit. Mass and spin of the BH are computed by means
of the isolated horizon formalism \cite{Ashtekar:2004cn,Dreyer02a} 
implemented in the \texttt{QuasiLocalMeasures} module.
In order to extract the GW signal, we decompose the Weyl scalar 
$\Psi_4$ into multipole moments on spheres of fixed coordinate 
radius. Computing the gravitational wave strain then requires 
twofold integration in time. In order to filter out the typical 
drift of the numerical waveform due to the cumulative effect of 
small numerical errors, we use the fixed frequency integration
technique described in \cite{Reisswig:2011}, with a cut frequency 
of $500\usk\hertz$. Since we are mainly interested in the GW 
spectra, we refrain from extrapolating the waveforms to infinity
and use a finite extraction radius 
near the outer boundary of the computational domain.

\subsection{Computing initial data}
\label{sec:initialdata}

The initial data for the irrotational binary models in this work
is computed using the publicly available \texttt{LORENE} 
code~\cite{Gourgoulhon:2000nn}.
To obtain properties of corresponding isolated, uniformly and
differentially rotating stars, we make use of the \texttt{RNS} 
code~\cite{Stergioulas95}.

In order to compute binaries with spin, we manually
add a rotational velocity field to irrotational models.
This is done as described in \cite{Kastaun2013},
by scaling the residual velocity field in the co-orbiting frame
according to
\begin{align}
\vec{\boldsymbol{w}} &= 
\left(1-s\right) \vec{\boldsymbol{w}}_L +
s\vec{\boldsymbol{\Omega}} \times \vec{\boldsymbol{x}}.
\end{align}
Above, $w^i = u^i/u^0$, $u^{\mu}$ is the fluid 4-velocity, 
$\vec{\boldsymbol{\Omega}}$ is the orbital angular velocity vector, 
$\vec{\boldsymbol{w}}_{_L}$ is the original irrotational velocity 
field, and $s$ is a free parameter that determines the amount 
of spin.

In contrast to the method described in \cite{Tichy11}, our
recipe for adding spin violates the constraint equations of 
GR. Thanks to the constraint-damping CCZ4 evolution scheme 
\cite{Alic:2011a}, the constraint violation becomes sufficiently 
small during the first millisecond of evolution, as detailed 
in \cite{Alic2013}. 
The constraint violations do however render the definition of
the initial angular momentum ambiguous. We use a definition 
of ADM angular momentum given by Eq.~(68) 
in~\cite{Gourgoulhon:2000nn}, which takes the form of a volume 
integral over fluid quantities, and which is very similar to 
the Newtonian expression.
Finally, we note that the modified rotational velocity profile 
does not satisfy hydrostatic equilibrium and thus introduces 
oscillations. For the spins considered here, the 
resulting oscillation amplitude is however less than 1\% in terms of the 
central density and can be safely ignored.

\subsection{Coordinates for analysis}
\label{sec:coordinates}

For our analysis of HMNS properties such as multipole
moments and rotation rates, we treat the system as a 
perturbation of a stationary, axisymmetric, and 
asymptotically flat background.
To this end, we require a more suitable coordinate system than 
the one used in our simulations. The main problem with 
the latter is that even when the system approaches an 
axisymmetric state after the merger, the spatial coordinates
do not reflect this, because the gauge conditions employed
in the evolution are not designed to seek symmetries.
In terms of proper distance axis ratios, 
the coordinate circles exhibit a deformation comparable with 
the physical deformation of the HMNS. Moreover, this deformation 
oscillates. The time slicing on the other hand is less 
problematic, judging by the lapse function which becomes 
reasonably stationary.

For those reasons, we construct a new coordinate system 
on each time slice in a postprocessing step. 
For simplicity, we restrict ourselves to the orbital plane
(the equatorial symmetry plane).
The new polar coordinates $(r,\phi)$
are defined in terms of the following requirements:
\begin{enumerate}
\item \label{enum:coord_req_origin}
The origin is the center of the $\pi$-symmetry of 
our models.
\item \label{enum:coord_req_grr}
Radial coordinate lines are parametrized by 
proper arc length, i.e., $g_{rr} = 1$.
\item \label{enum:coord_req_gpp}
The $\phi$-coordinate along constant $r$ is
proportional to proper arc length, i.e., $g_{\phi\phi,\phi}=0$.
\item \label{enum:coord_req_grp}
Along each circle $r=\text{const}$, the radial and polar 
coordinate lines are on average orthogonal, i.e.,
\begin{align}
  \int_{-\pi}^{\pi} g_{r\phi}(r,\phi) \,\mathrm{d}\phi &= 0\,.
\end{align}
\item \label{enum:coord_req_shift}
The shift vector $\beta^i$ of the new coordinates 
approaches zero for large radii.
\end{enumerate}

To show that such coordinates exist, one can treat the 
coordinate circles as a foliation of the orbital plane,
with a radial lapse function and a shift vector (not to 
be confused with the lapse and shift of the 3+1-decomposition).
Requirement \ref{enum:coord_req_grr} leads to an algebraic
expression for the lapse in terms of the shift.
Requirement \ref{enum:coord_req_gpp} yields an ordinary 
scalar differential equation of first order for the shift.
After taking the periodic boundary condition into account,
the solutions still possess one free parameter, which 
is fixed by requirement \ref{enum:coord_req_grp}.
Requirement \ref{enum:coord_req_shift} reduces the remaining 
freedom of a constant global rotation on each timeslice
to a rotation that is also constant in time, and which is 
irrelevant for our analysis.

To compute the above coordinates numerically, we use an 
iterative method acting on the whole coordinate system at once.
Each iteration consists of the following substeps, which 
each enforce one of the requirements, but preserve
the others only when the solution is reached:
\begin{enumerate}
\item
Compute the proper arclength of
radial lines, and use it as the new radial coordinate.
\item 
Compute the proper arclength 
along the coordinate circles
$l(r,\phi) = \int_0^\phi \mathrm{d}\phi' \sqrt{g_{\phi\phi}(r,\phi')}$ 
as well as the circumferential radius $2\pi \rcirc(r) = l(r, 2\pi)$
Using this, we then define a new $\phi$-coordinate 
$\phi \to l(r,\phi)/\rcirc(r)$.
\item 
Apply the mapping $\phi \to \phi + \delta\phi(r)$, where
$\delta\phi(r)$ fulfills the condition
\begin{align} \label{eq:shiftcorr}
  \frac{d}{d r} \delta\phi &= 
    -\frac{1}{2\pi r_c^2} 
    \int_{-\pi}^{\pi}  g_{r\phi} \, \mathrm{d}\phi\,.
\end{align}
\item
Apply a global rotation such that the $\phi=0$ coordinate 
is aligned with the $x$-axis at the outer edge of the computational 
domain.
\end{enumerate}
We stop iterating as soon as, \emph{during each substep}, 
the coordinates change less than a prescribed fraction 
(typically 0.1) of the grid resolution. As initial guess
for the iteration, we apply the canonical mapping to polar 
coordinates from the Cartesian simulation coordinates.
As a test, we mapped the Euclidean plane to strongly warped 
initial coordinates, and successfully recovered standard polar
coordinates. 

We use the new coordinates to define moments for any quantity
$q$ over a domain of radius $R$ by
\begin{align}\label{eq:moments}
P_m^q &= 
  \int_0^R  \int_0^{2\pi} q(r,\phi) e^{i m \phi} \mathrm{d}A\,,\\
p_m^q(r) &=
  \frac{1}{2\pi}
  \int_0^{2\pi} q(r,\phi) e^{i m \phi} \mathrm{d}\phi\,,
\end{align}
where the proper area form $\mathrm{d}A$ is given by
\begin{align}
\mathrm{d}A &=  
  \sqrt{g_{rr}g_{\phi\phi} - g_{r\phi}^2} \, \mathrm{d}r\, \mathrm{d}\phi
  &=
  \sqrt{r_c^2 - g_{r\phi}^2} \, \mathrm{d}r \, \mathrm{d}\phi\,.
\end{align}
For an axisymmetric manifold one can show that
$g_{r\phi}=0$, 
the radial coordinate $r$ is the Riemannian distance 
function\footnote{We note that using
the exponential map to construct coordinates is not 
a good choice in general. Even if the injectivity radius is larger than
the computational domain, which is not always the case,
$g_{\phi\phi}$ will vary strongly, which renders the
multipole moments meaningless.},
and the radial coordinate lines are geodesics. 
Together with the condition $g_{\phi\phi,\phi}=0$,
we find that $\partial_\phi$ is a Killing vector.
Hence, \emph{for solutions axisymmetric around the origin, 
all moments $P_m$ and $p_m(r)$ of physical quantities 
vanish unless $m=0$}. Naturally, the moments above are
not useful if the NS is not centered around the coordinate 
origin. For this reason, we do not use the framework 
for unequal-mass models in this work.

Next, we define a fluid angular velocity by
\begin{align}\label{eq:omega_rot}
\Omega(t,r,\phi) &= 
\frac{\partial \phi}{\partial x^b}\left( 
  \alpha v^b - \bar{\beta}^b
\right).
\end{align}
Here, $\alpha$ is the lapse function, 
$x^b$ denotes the simulation coordinates,
and $v^b$ the fluid 3-velocity.
$\bar{\beta}^b$ is the shift vector corresponding to the new 
coordinates, related to the shift vector $\beta^b$ 
of the simulation coordinates by
\begin{align}
\bar{\beta}^b &= 
  \beta^b + \frac{\partial \bar{x}^b}{\partial t},
\end{align}
where $\bar{x}^b$ are the new coordinates in terms of the old 
ones.
We note that in the limit of a stationary and axisymmetric 
spacetime, $\bar{\beta}^\phi$ is a measure of the frame dragging,
and $\Omega$ is the angular frequency as observed from infinity.
Further, rigid rotation is equivalent to $\Omega = \text{const}$.
Also note that even in the limit of large radii, points of constant 
$r,\phi$ can be subject to acceleration, because the proper
distance to the origin also depends on the geometry in the 
strong field region. For this reason, we prefer to use the 
circumferential radius $\rcirc$, which depends on the geometry 
at smaller radii less directly, via the shape of coordinate 
spheres.

\begin{figure}
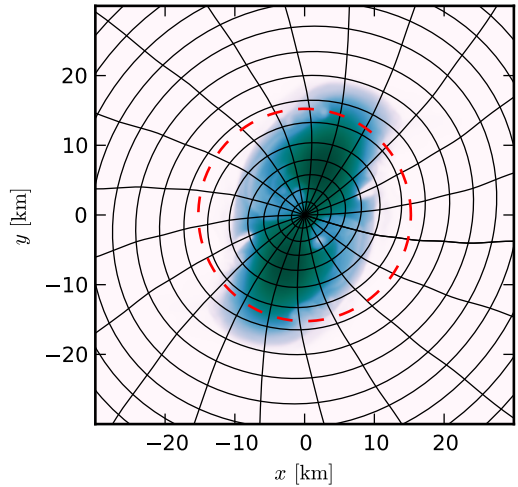

  \centering
  \includegraphics[width=0.99\columnwidth]{{{new_coords_xy_SHT-M2.0-I}}}
  \caption{The polar coordinate system constructed for analysis
  at a time $0.9\usk\milli\second$ after the merger, for model 
  \texttt{SHT-M2.0-I}. The black radial 
  lines are the coordinate lines of constant $\phi$, and the black 
  circles mark coordinate lines of constant $\rcirc$.
  Both are spaced equidistantly. For comparison, we show the density 
  profile as color plot, and a coordinate circle with respect to 
  simulation coordinates (dashed red curve).} 
  \label{fig:polar_coords}
\end{figure}

\begin{figure}
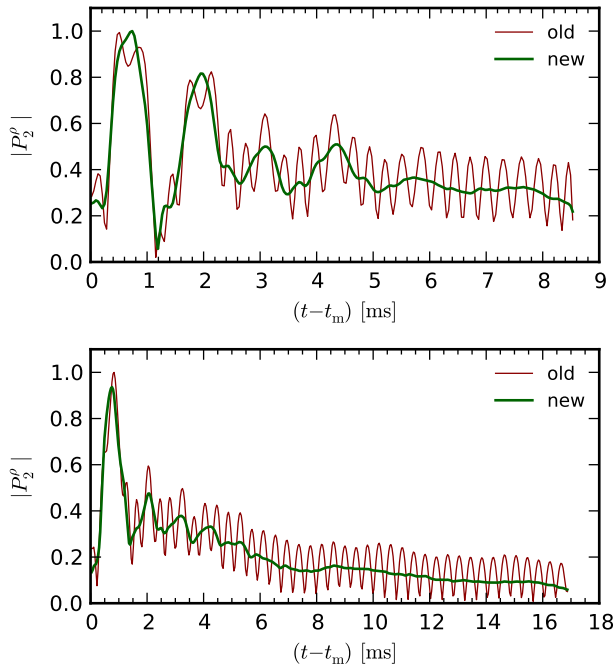

  \centering
  \includegraphics[width=0.99\columnwidth]{{{compare_moments_LS220-M1.5-I}}}
    \includegraphics[width=0.99\columnwidth]{{{compare_moments_SHT-M2.0-I}}}
  \caption{Comparison of density moments $P_2^\rho$ computed 
  either using the new coordinate system (thick green line) or the 
  simulation coordinates (thin red line). Shown are the results for for 
  models \texttt{LS220-M1.5-I} (top panel) and \texttt{SHT-M2.0-I} 
  (bottom panel).
  } 
  \label{fig:compare_coords_moment}
\end{figure}

Figure~\ref{fig:polar_coords} shows the new coordinates for one of our
simulations shortly after the merger. In comparison, the simulation 
coordinates exhibit a dominant $m=2$ distortion as well as a 
spiral deformation. 
The deformation remains mostly stationary after the merger, with 
a slight wobbling synchronized to the oscillations of the HMNS. 
At late times, the physical deformations of the HMNS we want 
to measure become smaller than the coordinate deformation.
For the models considered in this work, the deformations
of the coordinates mainly affect the $m=2$ and $m=4$ moment
amplitudes and phase velocities. 
This can be seen in \Fref{fig:compare_coords_moment}, where we 
compare the amplitude $|P_2^\rho|$ computed using the new 
coordinates to values obtained using the canonic polar coordinates 
associated with the Cartesian simulation coordinates. The (complex 
valued) moment acquires a constant offset when using the simulation 
coordinates, which leads to a modulation of the moments absolute 
value even if the physical amplitude of the NS oscillation remains 
constant.
For $m=0$ moments like average rotation rate or radius, the 
differences turn out to be minor in our simulations. 
Still, the new coordinates are useful to exclude gauge effects.
We conclude that the transformation to well defined coordinates is
essential to get reliable values for multipole moments of HMNSs
and suppress gauge artifacts.

\section{Models}
\label{sec:models}

For the simulations presented in this work, we consider two 
different nuclear physics EOSs which incorporate thermal effects  
and composition (electron fraction). 
One is the Lattimer-Swesty (LS) EOS \cite{Lattimer91}, which 
is based on a compressible liquid drop model with Skyrme interaction
\cite{Tondeur1984}. We use the variant with incompressibility 
modulus $K = 220 \usk\mega\electronvolt$.
The other one is the Shen-Horowitz-Teige (SHT) EOS
(see \cite{ShenG2010,ShenG2011}),
which is derived from a relativistic mean-field model for 
uniform matter with a modified NL3 set of interaction parameters. 
To construct initial data, we assume cold matter 
at $\beta$-equilibrium.

We study several irrotational equal-mass binaries 
with different total mass. 
In addition, we evolve two of the equal-mass binaries with 
NS spins aligned to the orbital angular momentum.
We also report results for one unequal-mass irrotational 
binary (we note this model was still evolved using the BSSNOK 
instead of the CCZ4 formulation).
All our models are summarized in \Tref{tab:models}.

To quantify the NS spin, we compute the change $\Delta \Jadm$ 
of total ADM angular momentum (see also \cite{Kastaun2013}) 
of the binary with respect to the 
irrotational model of same baryonic mass and separation.
The NS spins in \Tref{tab:models} are given in terms of
the frequency
$2 \pi \Delta F_\mathrm{R} = \frac{1}{2}\Delta \Jadm / I_\infty$, 
where $I_\infty$ is the moment of inertia of the nonrotating 
NS in isolation.
Compared to the rotation rate of $44.05 \usk\hertz$ observed 
for the well known double pulsar PSR J0737-3039 A, 
the chosen values for the spinup $\Delta F_\mathrm{R}$ 
are larger, but they are still small compared to the Kepler 
limit. The latter is reached at a rotation rate of
$893\usk\hertz$ for models \texttt{LS220-M1.5-I} and~\texttt{-S}, 
and $814\usk\hertz$ for models \texttt{SHT-M2.0-I} and~\texttt{-S}.
The spinup corresponds to an increase $\approx 0.1$ 
of the dimensionless spin 
$\Delta \NSspin = \frac{1}{2} \Delta \Jadm / M_\infty^2$,
where $M_\infty$ is the gravitational mass of the NS in isolation.

The initial separations are large enough to allow 6--11 orbits
until merger. The total baryonic mass is chosen around the threshold mass
for prompt BH collapse given in \cite{Bauswein2013}, since 
we are mainly interested in models which lead to a stable HMNS 
or delayed collapse.
For comparison, the maximum baryonic mass of cold nonrotating 
stars is 
$\Mbar = 2.42 \usk M_\odot$ for the LS EOS 
and 
$\Mbar = 3.38 \usk M_\odot$ for the SHT EOS. 
The maximum mass of rigidly rotating stars is 
$\Mbar = 2.83 \usk M_\odot$ 
and 
$\Mbar = 3.97 \usk M_\odot$, respectively.

The setup of the numerical grid is identical for spinning and 
irrotational models, and the finest grid resolution is 
$295\usk\meter$ for all models.
The outer boundaries of the numerical domain are located a 
$803 \usk\kilo\meter$ for the LS220 models, 
and $945 \usk\kilo\meter$ for the SHT models.

\begin{table}
\begin{ruledtabular}
\begin{tabular}{lccccc}
Model&
$\Mbar [\Msol]$&
$M_\infty [\Msol]$&
$\Delta F_\mathrm{R} [\hertz]$&
$\Omega_\mathrm{o} [\radian\per\second]$&
$d [\kilo\meter]$
\\\hline 
SHT-M2.0-I&
$4.01$&
$1.80$&
$0$&
$1460$&
$74.2$
\\
SHT-M2.0-S&
$4.01$&
$1.80$&
$155.4$&
$1460$&
$74.2$
\\
SHT-M2.2-I&
$4.39$&
$1.95$&
$0$&
$1505$&
$76.0$
\\
LS220-M1.5-I&
$3.12$&
$1.41$&
$0$&
$1552$&
$64.5$
\\
LS220-M1.5-S&
$3.12$&
$1.41$&
$159.8$&
$1552$&
$64.5$
\\
LS220-M1.7-I&
$3.46$&
$1.54$&
$0$&
$1611$&
$66.2$
\\
LS220-M1.8-I&
$3.62$&
$1.61$&
$0$&
$1636$&
$66.9$
\\
LS220-M1.4-U&
$2.94$&
$1.26,1.41$&
$0$&
$1520$&
$63.3$
\end{tabular}
\end{ruledtabular}
\caption{Parameters of the binary NS models. 
$\Mbar$ is the total baryonic mass of the binary
(Baryonic masses are based on a formal baryon mass of 
$m_\mathrm{B}=931.494 \usk\mega\electronvolt$ throughout this 
work).
$M_\infty$ is the gravitational mass of each star at
infinite separation. $\Omega_\mathrm{o}$ is the initial 
orbital angular velocity, and $d$ the initial
proper distance between density maxima locations.
$\Delta F_\mathrm{R}$ is the NSs rotation rate compared
to the irrotational case (see main text).}
\label{tab:models}
\end{table}

\section{Results}
\label{sec:results}
In this section we present the outcome of our numerical 
simulations.
For a discussion of the numerical accuracy of spinning BNS simulations
using the same code, we refer the reader to \cite{Alic2013}, while the 
particularities of using tabulated EOSs are discussed in 
\cite{Galeazzi2013}.

\subsection{Inspiral phase and merger outcome}
\label{sec:inspiral}

\begin{figure}
  \centering
  \includegraphics[width=0.99\columnwidth]{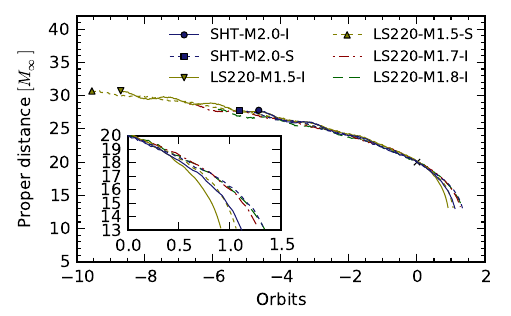}
  \caption{Proper separation of maximum density locations versus
  their orbital phase. The phase has been aligned at a fixed
  separation of $d=20 \usk M_\infty$. The symbols are marking 
  the start of the simulations. }
  \label{fig:sep_vs_phase}
\end{figure}

To measure the orbital trajectories, we use the positions of
the rest mass density maximum with respect to the coordinates
defined in \Sref{sec:coordinates}. 
The proper separation for the equal-mass models is estimated
from the radial coordinate (which equals the proper distance
along radial coordinate lines), and the orbital phase is defined
as the $\phi$-coordinate.
Figure~\ref{fig:sep_vs_phase} shows the proper distance versus 
the orbital phase. 
To compare different models, we align the phase at a separation 
of $d=20 \usk M_\infty$, where finite size effects should still 
be small. In units of the NS gravitational mass, the separations 
are indeed very similar up to this point, as
it would be the case for point particles.
The differences are comparable to the eccentricity caused 
by the inaccuracies of the initial data.
For smaller separations, shown in the 
inset of \Fref{fig:sep_vs_phase}, the discrepancies
between the models increase, as finite size effects 
become more and more significant.
We did not try to relate the orbital 
trajectories to models of tidal effects. 
However, we do observe an influence of the NS spin.
When comparing spinning and irrotational simulations
with otherwise identical setup, in particular 
starting at the same proper separation, we find that
the spinning models require more orbits to reach a given 
separation.
The influence of the spin grows stronger closer to the 
merger, as visible in the inset of \Fref{fig:sep_vs_phase}.
This "orbital hangup" has already been observed in our
previous work \cite{Kastaun2013} for models with an ideal gas EOS, 
as well as in \cite{Tsatsin2013,Bernuzzi2013prd} using two different methods
of adding the spin.

The main properties of the merger remnants are summarized 
in \Tref{tab:mergers}.
To quantify the HMNS lifetime, we define
$\tau_\text{HMNS} = t_\text{BH} - t_\text{m}$, where
$t_\text{BH}$ is the (coordinate) time the apparent horizon 
first appears, and $t_\mathrm{m}$ is the time of the first 
maximum of the rest mass density after the stars touch. 
If the maximum density just increases until a BH is formed 
we call it a prompt collapse.
The values reported for $\tau_\text{HMNS}$ should be taken 
with care, since HMNS lifetimes longer than a few 
$\milli\second$ are increasingly sensitive to numerical 
errors. However, we can reliably determine the minimum 
mass $\Minf$ required for prompt collapse, $\Minf^\text{thr}$.
From Tables~\ref{tab:models} and ~\ref{tab:mergers}, we 
obtain a value
$\Minf^\text{thr} = 1.5 \usk\Msol$ for the LS EOS, and
$\Minf^\text{thr} = 1.9 \usk\Msol$ for the SHT EOS,
with an estimated error of $\pm 0.1 \usk\Msol$.
This confirms in full GR
the respective values reported in \cite{Bauswein2013}, 
which were obtained using the conformal flatness 
approximation.

\begin{figure}
  \centering
  \includegraphics[width=0.99\columnwidth]{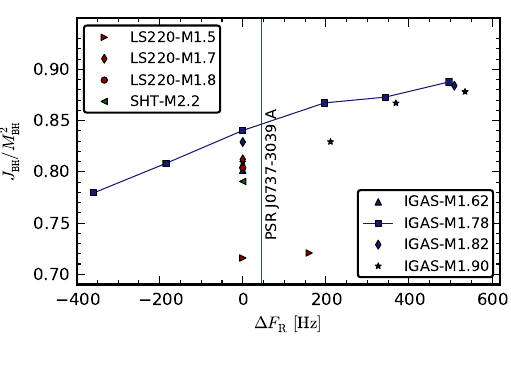}
  \caption{Dimensionless BH spin versus initial NS spinup
  rate $\Delta F_\mathrm{R}$ (see main text).
  Our results are shown in comparison to earlier results 
  \cite{Kastaun2013} for the case of an ideal gas EOS. 
  The vertical line marks the rotation rate of the double 
  pulsar PSR J0737-3039A.
  }
  \label{fig:bh_vs_ns_spins}
\end{figure}

We now discuss the spin of the final BH, $\BHspin=J_{\rm{BH}}/M^2_{\rm{BH}}$, 
given in \Tref{tab:mergers}. In \Fref{fig:bh_vs_ns_spins}, it is shown as a function
of the additional NS spin,
together with earlier results from \cite{Kastaun2013} for 
an ideal gas EOS.
Again, we find a slightly increased BH spin for the spinning 
model compared to the irrotational model. The absolute BH spin 
for models \texttt{LS220-M1.5-I} and~\texttt{-S} (the ones 
resulting in a delayed collapse) is however significantly lower 
than for the other models, in particular the heavier ones with 
identical EOS. 
The disk remaining outside the BH on the other hand
possesses ${\approx}9$\% of the BH angular momentum,
while for the heavier models, mass and angular momentum
of the disks are negligible.
We conclude that the main reason for the lower BH spin
is the angular momentum contained in the disk surrounding the 
HMNS.

\begin{table}
\begin{ruledtabular}
\begin{tabular}{lcccc}
Model&
$\tau_\text{HMNS}\, [\milli\second]$&
$\Mbh \, [\Msol]$&
$\BHspin$&
$\Mdisk\,[\Msol]$
\\\hline 
SHT-M2.0-I&
$>16.9$&
---&
---&
---
\\
SHT-M2.0-S&
$>9.4$&
---&
---&
---
\\
SHT-M2.2-I&
prompt&
$3.73$&
$0.791$&
$0.00$
\\
LS220-M1.5-I&
$8.6$&
$2.65$&
$0.716$&
$0.06$
\\
LS220-M1.5-S&
$7.7$&
$2.67$&
$0.721$&
$0.04$
\\
LS220-M1.7-I&
prompt&
$2.98$&
$0.812$&
$0.00$
\\
LS220-M1.8-I&
prompt&
$3.14$&
$0.804$&
$0.00$
\\
LS220-M1.4-U&
$>29.8$&
---&
---&
---
\end{tabular}
\end{ruledtabular}
\caption{Key properties of the mergers. 
$\tau_\text{HMNS}$ denotes the HMNS lifetime (see main text),
unless the merger results in a prompt collapse to a BH.
For models which did not collapse we report the post-merger 
duration of the simulation instead. 
$\Mbh$ is the final BH mass, $\BHspin = \Jbh / \Mbh^2$ its
dimensionless spin. The baryonic mass outside the BH at the
end of each simulation is denoted by $\Mdisk$.}
\label{tab:mergers}
\end{table}

\subsection{Hypermassive neutron star properties}
\label{sec:hmns}

In this section we investigate the properties and dynamics 
of the HMNSs. For this purpose, we use the moments
in the orbital plane defined by \Eref{eq:moments},
with a cutoff radius $R=30\usk\kilo\meter$ that is
sufficiently large to cover the oscillating HMNS.
We recall that the construction of the coordinates used in 
the definition of the moments eliminates ambiguities due 
to the spatial gauge of the coordinates used in the simulation.

\begin{figure}
  \centering
  \includegraphics[width=0.99\columnwidth]{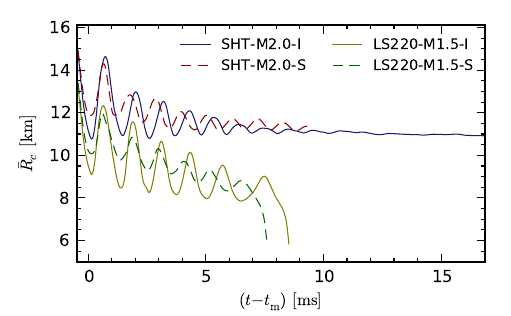}
  \caption{Evolution of the HMNS average circumferential
  radius $\bar{R}_c$. The time coordinate is relative to
  the merger time $t_m$.}
  \label{fig:hmns_radius_evol}
\end{figure}

The most basic HMNS property is the size,
which we quantify in terms of the density weighted 
average circumferential radius, i.e.,
$\bar{R}_c = P_0^{\rho\, r_c} / P_0^\rho$.
The time evolution of this measure is shown in 
\Fref{fig:hmns_radius_evol}, while the corresponding
oscillation frequencies are reported in \Tref{tab:hmns_freqs}.
For the irrotational model \texttt{SHT-M2.0-I}, the star initially 
undergoes strong radial oscillations, which are however
damped almost completely within $10\usk\milli\second$
after the merger. The stellar radius $\bar{R}_c$ changes
only very slightly after this point until the end of the 
simulation. Therefore, this model might remain stable 
considerably longer. 

The evolution of $\bar{R}_c$ for spinning model 
\texttt{SHT-M2.0-S} differs mainly in one respect from the irrotational
case. As one can see in \Fref{fig:hmns_radius_evol}, 
the oscillation amplitude as
well as the maximum compactness reached for the spinning 
case are smaller.
The reason for this effect might be a larger centrifugal barrier
due to the additional angular momentum. On the other hand, the 
compactness differs much less once the stars settle down.
Equally likely, the difference may be caused by the
slightly different impact trajectories, i.e., the orbital 
hangup discussed in \Sref{sec:inspiral}.
The different oscillation amplitude will turn out to be 
important for the GW spectrum and for the ejection of matter,
as we show below.

The HMNSs for models \texttt{LS220-M1.5-I} and \texttt{-S},
although less massive, are smaller than the 
\texttt{SHT-M2.0} models.
Their radii show a continuous 
drift in addition to the oscillations, until they collapse 
to a BH. When approaching the collapse, the frequency of 
the radial oscillation strongly decreases. This is to be 
expected, since the frequency of the 
(linear) quasi-radial mode also approaches zero for 
marginally stable models of isolated NSs. 
The radial frequency given in \Tref{tab:hmns_freqs} in 
this case represents only an average value.
We note that also for 
those models, the irrotational one becomes more compact
than the spinning one during the initial oscillations.

\begin{figure}
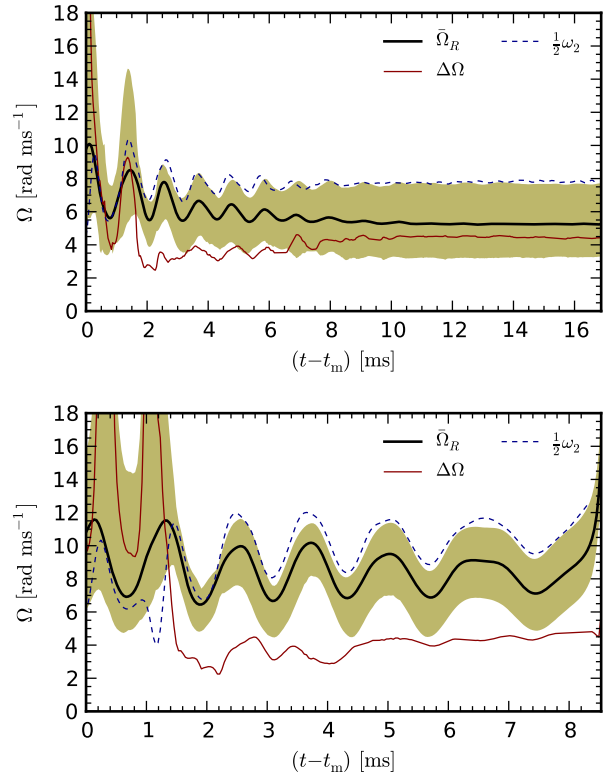

  \centering
  \includegraphics[width=0.99\columnwidth]{{{hmns_moments_freqs_SHT-M2.0-I}}}\\
  \includegraphics[width=0.99\columnwidth]{{{hmns_moments_freqs_LS220-M1.5-I}}}
  \caption{Time evolution of the HMNS rotation for models
  \texttt{SHT-M2.0-I} (top panel), and \texttt{LS220-M1.5-I} (lower panel). 
  The black solid line shows the average angular velocity
  $\bar{\Omega}_\mathrm{R}$. 
  The shaded area is bounded by the extrema of the angular velocity 
  $\omega_\mathrm{R}$ in the orbital plane inside the HMNS (ignoring 
  mass densities below $10^{-2}$ of the maximum one).
  The difference $\Delta\Omega$ between maximum and minimum is given by
  the red curve.
  The dotted line shows the pattern angular velocity 
  $\frac{1}{2}\omega_2$
  of the $l=m=2$ mode, computed from the phase velocity of 
  $P_2^\rho$. }
  \label{fig:hmns_rot_evol}
\end{figure}

Next, we study the rotation of the HMNSs. We define a density
weighted average angular velocity in the orbital plane as
$\bar{\Omega}_\mathrm{R} = P_0^{\rho\Omega} / P_0^{\rho}$,
where $\Omega$ is defined by \Eref{eq:omega_rot}.
\Fref{fig:hmns_rot_evol} depicts the evolution of the 
rotation rate. For all models, it shows a strong modulation, 
which turns out to be anti-correlated with the radius $\bar{R}_c$.
This should be expected if angular momentum can be considered 
constant on the timescale of the oscillation period.
Note that the radial oscillation frequency 
is similar to the rotation rate for models \texttt{SHT-M2.0-I} 
and~\texttt{-S}, and even smaller for models \texttt{LS220-M1.5-I} 
and~\texttt{-S}.

To quantify the amount of differential rotation,
we use the $\phi$-averaged rotation rate 
$\omega_\mathrm{R}(r) = p_0^\Omega(r)$.
We recall that rigid rotation implies that
$\Omega$, and hence $\omega_\mathrm{R}$, are constant.
We compute the extrema of $\omega_\mathrm{R}$ within the 
HMNS at each time, shown in \Fref{fig:hmns_rot_evol}.
The average values are reported in \Tref{tab:hmns_freqs}.
All our HMNSs possess a high degree of differential 
rotation, which fluctuates less than the rotation rate 
itself. 
Interestingly, it is slowly increasing. 
This has to be a consequence of the change
in compactness, since dissipative effects can only 
decrease the differential rotation.

\begin{figure}
  \centering
  \includegraphics[width=0.99\columnwidth]{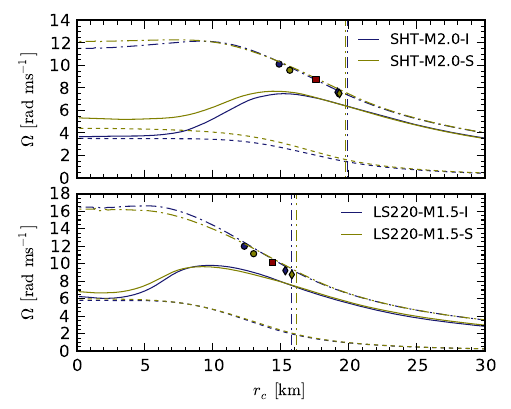}
  \caption{Average rotation profile and Keplerian angular 
  velocity profile for models \texttt{SHT-M2.0-I} and~\texttt{-S} 
  (top panel), and \texttt{LS220-M1.5-I} and~\texttt{-S} 
  (bottom panel).  
  The solid curves mark the angular velocity $\omega_\mathrm{R}$ 
  in the orbital plane versus circumferential radius $r_c$. 
  The dotted lines show the contribution $-\bar{\beta}^\phi$ 
  of the frame dragging.
  The dash-dotted curves are an estimate of the angular velocity
  of test particles in corotating circular orbits.
  The vertical dashdotted lines mark the radius where the 
  mass density falls below 5\% of the central one.
  For comparison, the circles mark radius and equatorial orbital velocity
  of cold, uniformly rotating stars with the same central rotation rate
  and central rest mass density.
  The diamonds mark models at zero-temperature
  with the same central rest mass density, uniformly rotating
  at Keplerian rate.
  The red squares mark the maximum mass (Keplerian) model for 
  uniformly rotating cold NSs.
  }
  \label{fig:hmns_rot_prof}
\end{figure}

The rotation profile $\omega_\mathrm{R}(r)$ reached after the 
stars have settled down is pictured in \Fref{fig:hmns_rot_prof}. 
In order to eliminate the modulation due to the oscillations,
we construct an averaged neutron star (ANS in the following),
averaging first in $\phi$-direction and then in time.
The time interval chosen for averaging is
$5$--$9 \usk\milli\second$ after the merger for the SHT models, 
and $3$--$7 \usk\milli\second$ for the LS220 models.
The rotation rate at the center is close to the minimum one, while
the maximum appears in the outer layers of the star.
This is true at all times, except during the merger phase,
where the maximum rotation rate is reached in the central region.
When comparing the spinning models to the irrotational ones,
we find an increased average rotation rate, as one would expect.
The maximum rotation rate of the spinning models
on the other hand is similar or even smaller than for
the irrotational models.

On a side note, the rotation as seen from infinity in the 
center of the ANSs is almost completely due to the frame 
dragging effect, i.e., the fluid is rotating slowly with 
respect to the local inertial frame. 
This is visualized in \Fref{fig:hmns_rot_prof}, which also
contains the frame dragging contribution $-\bar{\beta}^\phi$ to 
the rotation rate $\Omega$, see \Eref{eq:omega_rot}. 

\begin{figure}
  \centering
  \includegraphics[width=0.99\columnwidth]{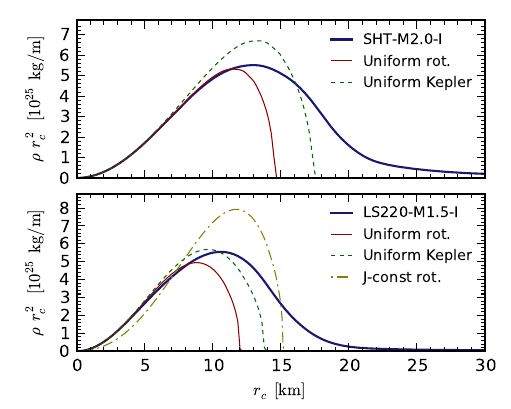}
  \caption{Density profile in the equatorial plane of
  the time and $\phi$-averaged HMNS for models 
  \texttt{SHT-M2.0-I} (top panel), 
  and \texttt{LS220-M1.5-I} (bottom panel).
  For comparison, we show the density profiles of uniformly rotating
  NS stars with same central density and central rotation rate as the 
  HMNS, and same central density but maximum rotation rate.
  For the LS220 case, we also show a model with same mass as the HMNS, 
  but a $j$-const rotation law (see main text).
  The density is scaled by $r_c^2$ to better visualize the contribution 
  of each  spherical shell to the total mass. Note however that since the
  models are not spherically symmetric, simply taking the integrals of 
  the curves would overestimate the total mass inside a given radius.
  }
  \label{fig:hmns_dens_prof}
\end{figure}

Armed with the rotation profile, we will now investigate why 
the system does not immediately collapse to a BH
despite
the total baryonic mass exceeding the maximum one for a 
cold, uniformly rotating star in $\beta$-equilibrium. 
There are two possibilities: thermal effects and differential 
rotation. As pointed out in \cite{Kaplan2013apj}, 
increasing the (uniform) temperature tends to decrease 
the maximum mass of stars rotating uniformly or 
differentially with a $j$-constant law.
In \cite{Galeazzi2013}, the same was found for non-rotating
stars with constant specific entropy.
Further, we find that the maximum baryonic 
mass of uniformly rotating stars obeying the LS220 EOS with 
constant specific entropy of $4\usk k_\mathrm{B}$ (and 
$\beta$-equilibrium) is reduced by $15$\% compared to the 
zero-temperature case. Although the situation for differential
rotation may be different, it seems unlikely that 
temperature is responsible for stabilizing the merger remnants
produced in our simulations.

\begin{table}
\begin{ruledtabular}
\begin{tabular}{lcccc}
Model&
$F_\mathrm{Rot}^\mathrm{min}\, [\kilo\hertz]$&
$F_\mathrm{Rot}^\mathrm{max}\, [\kilo\hertz]$&
$F_\mathrm{Rad}\, [\kilo\hertz]$&
$F_2^2\, [\kilo\hertz]$
\\\hline 
SHT-M2.0-I&
$0.66$&
$1.20$&
$0.94$&
$2.47$
\\
SHT-M2.0-S&
$0.85$&
$1.45$&
$0.95$&
$2.66$
\\
LS220-M1.5-I&
$0.96$&
$1.56$&
$0.76$&
$3.24$
\\
LS220-M1.5-S&
$1.05$&
$1.62$&
$0.89$&
$3.17$
\\
LS220-M1.4-U&
n.a.&
n.a.&
$1.45$&
$2.93$
\end{tabular}
\end{ruledtabular}
\caption{HMNS frequencies. The minimum and maximum rotation
  rates inside the HMNS are denoted
  $F_\mathrm{Rot}^\mathrm{min}$ and $F_\mathrm{Rot}^\mathrm{max}$,
  respectively. The values are time averages over the interval
  starting $2\usk\milli\second$ after the merger until the end of 
  the simulation or the collapse to a BH.
  The radial oscillation frequency $F_\mathrm{Rad}$ was computed
  from the Fourier spectrum of $\bar{R}_c$. 
  The average frequency $F_2^2$ of the $l=m=2$ oscillation mode 
  was extracted from GW effective strain spectrum.}
\label{tab:hmns_freqs}
\end{table}

We are left with differential rotation to support larger
masses than the maximally but uniformly rotating models.
This has been 
discussed mainly for rotation laws where the core
rotates faster than the outer layers, 
see \cite{Baumgarte00bb,Kaplan2013apj}.
However, 
this picture does not apply to the remnants formed in our 
simulations. We computed the stellar models with the same 
central rest mass density as the ANS, but rotating uniformly 
at the Kepler limit, as well as the maximum mass Kepler model.
Radius and rotation rate are compared to the ANS rotation 
profile in \Fref{fig:hmns_rot_prof}.
As one can see, the rotation rate of the ANS core is 
significantly lower. 
The notion that  a dense core rotating more rapidly than the
Keplerian model is responsible for stabilizing the mass 
against collapse is thus ruled out for the equal-mass models 
presented here (but not in general).
Note a similar rotation profile has also
been reported in Fig.~8 of~\cite{Shibata05c} for different 
models. 
Besides the rotation rate, the central density of the ANS 
is also lower than the one of the Keplerian maximum mass 
model, around $75\%$ and $83\%$ for the
for the SHT and LS220 EOS models, respectively.

As pointed out in \cite{Baumgarte00bb}, another option to 
increase the maximum mass is the addition of an extended 
Keplerian disk. Even if the rotation profile is only close 
to the Keplerian one, it should be possible to create a massive 
envelope. 
Indeed, we find that the density profile of our HMNS extends
to larger radii than the ones of uniformly rotating stars.
As shown in \Fref{fig:hmns_dens_prof}, the density profiles of 
uniformly rotating stars with same central density agree well
in the core region, but the HMNS extends well beyond the surface
of the uniformly rotating stars. 
For comparison, the baryonic masses of the uniformly rotating 
models with same central density and same central rotation rate 
(maximum rotation rate) are
$2.46 \usk\Msol$ ($2.78 \usk\Msol$) and 
$3.12 \usk\Msol$ ($3.74 \usk\Msol$) for 
the LS220 and SHT models, respectively. This is $7$ -- $22\%$
lighter than the total masses in our simulations (compare \Tref{tab:models}).
Figure~\ref{fig:hmns_dens_prof} seems to indicate that the mass in the 
core regions of the HMNSs, up to ${\approx}8\usk\kilo\meter$ 
for the LS220 models and ${\approx}12\usk\kilo\meter$ for the 
SHT models, is similar to the uniform models, while supporting 
more mass further out. 
Note however that the density profile in the equatorial plane
can only provide a very rough estimate for total mass in a given 
radius, due to the deviation from spherical symmetry. For lack of 
saved 3D data, we could not compare the exact mass-inside-radius 
relations. 

Without doubt, there is a significant amount of mass in the extended 
outer layers. It is therefore natural to ask how close
to Keplerian velocity our profiles are at a given radius.
To obtain an estimate, we approximate the spacetime as stationary 
and axisymmetric with respect to the coordinates introduced in 
\Sref{sec:coordinates}. We compute the metric components in the 
orbital plane using the same averaging as for the ANS rotation 
profile.
The angular velocity $\Omega_\mathrm{K}$ of test particles
in a corotating circular orbit at each radius is then given by 
the larger solution of the quadratic equation
\begin{align}
0 &= g_{tt,r} + 2 g_{t\phi,r} \Omega_\mathrm{K} 
     + g_{\phi\phi,r} \Omega_\mathrm{K}^2\,.
\end{align}
The result, shown in \Fref{fig:hmns_rot_prof}, is that the ANSs
have a large envelope moving shortly below Keplerian velocity. 
However, it is difficult to estimate the error of the Keplerian 
velocity profile due to the non-stationarity of the actual spacetime.
As a cross-check, we also compare with the equatorial orbital 
velocity of uniformly rotating models with the same central 
density and rotation rate as the ANSs (circles in \Fref{fig:hmns_rot_prof}).
Assuming that the orbital velocities are mainly determined
by the dense core, this provides us with a rough, but independent 
estimate. As shown in \Fref{fig:hmns_rot_prof}, it matches the first 
one surprisingly well. Curiously, also the maximum mass Kepler models 
(red squares in \Fref{fig:hmns_rot_prof}) are located
very close to the Kepler profile of the simulation results.

We note that the rotation profile is only an average of
a very dynamic system, in particular for the LS220 models,
which also undergo a drift to higher compactness in addition
to the large oscillations.
That said, \emph{the above findings strongly suggest that the system
is stabilized against immediate collapse by an extended, almost-Keplerian 
envelope}.
We consider this to be the important aspect of the rotation 
profile with respect to stabilizing the mass, not the slow 
rotation of the core. 
For the irrotational LS220 model, we checked that it is also possible 
to construct a HMNS with the same mass but a $j$-const rotation
law. The model we constructed is shown in 
\Fref{fig:hmns_dens_prof}. It has a central rotation rate of 
$\Omega_c=1.694 \usk\kilo\hertz$ and a differential rotation parameter 
$A=1.48\usk\kilo\meter$.
In contrast to the other models, the density maximum is not at 
the center due to the rapid core rotation.

\begin{figure}
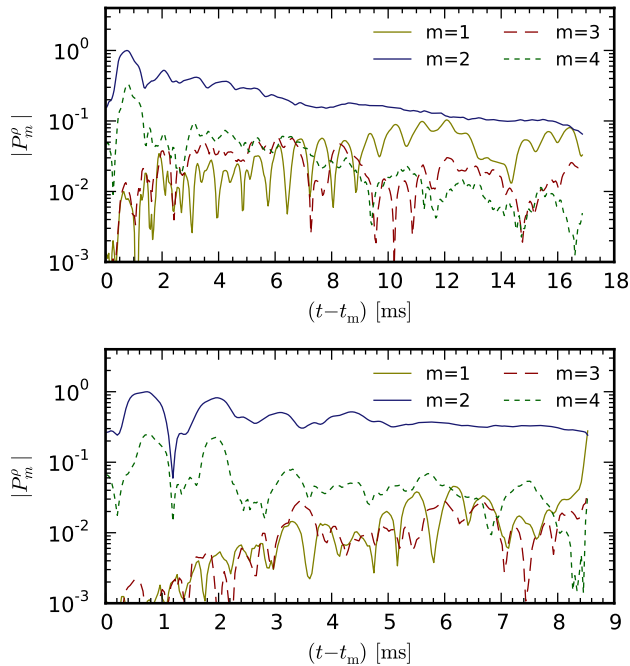

  \centering
  \includegraphics[width=0.99\columnwidth]{{{hmns_moments_amp_SHT-M2.0-I}}}\\
  \includegraphics[width=0.99\columnwidth]{{{hmns_moments_amp_LS220-M1.5-I}}}
  \caption{Time evolution of the density moment amplitudes 
  $|P^\rho_m|$ for models \texttt{SHT-M2.0-I} (top panel) and
  \texttt{LS220-M1.5-I} (bottom panel). The amplitudes are normalized
  to the maximum of the $m=2$ moment.}
  \label{fig:hmns_moments_amp}
\end{figure}

We now turn our attention to the non-axisymmetric perturbations 
of the HMNS. 
The amplitudes of the density moments $P_m^\rho$ are plotted in 
\Fref{fig:hmns_moments_amp}. We first discuss the $m=2$ component,
which is clearly dominant. Obviously, it is excited
by the merger and then slowly decays.
In order to interpret the meaning of $P_2^\rho(t)$, we also 
studied the phase of $p_2^\rho(t, r)$. 
We found that during the plunge, the phase of $p_2^\rho$ 
naturally follows the average orbital phase of the two stellar 
cores.
During the subsequent merger stage the phase varies strongly 
inside the forming HMNS. This could be caused by the presence
of several oscillations of comparable strength. However,
we feel that this stage is genuine nonlinear
and do not interpret it in terms of perturbation theory.
1--2$\usk\milli\second$ after the merger 
time $t_m$ however, the $m=2$ perturbation becomes dominated 
by a single oscillation mode, i.e., the phase of $p_2^\rho$ 
is independent of the radius.
Only in the outer layers of the star, at densities around $5\%$
of the central one, the oscillations are out of phase. 
Considering the large oscillation amplitude, this is quite 
normal. As discussed in \cite{Kastaun2010}, the nonlinear 
effects at the surface will contribute to the 
damping of the mode, probably more than 
gravitational radiation.

Since during the HMNS stage there exists one strongly dominant 
$m=2$ oscillation mode, the phase velocity of $P_2^\rho$ is
a good estimate for its instantaneous oscillation frequency
$\omega_2$. In \Fref{fig:hmns_rot_evol}, 
$\omega_2$ is shown in comparison to the rotation 
rates. Note the value is only meaningful once $P_2^\rho$ is
dominated by the $m=2$ oscillation mode, which happens after
$\approx2\usk\milli\second$. As one can see,
\emph{the frequency is strongly modulated. This has to
be taken into account when interpreting the Fourier spectra of GW 
signals}, as will be discussed in \Sref{sec:gw}.
Clearly, the frequency is also strongly correlated with the 
rotation rate. A heuristic explanation can be obtained by regarding 
the radially oscillating HMNS as a sequence of fixed background 
models for the $m=2$ mode oscillation, which has a higher frequency.
For uniformly rotating stars, it is known that the frequency 
in the corotating frame depends weakly on rotation rate. 
The frequency of the corotating mode in the 
inertial frame is shifted by $+m\Omega$. The resulting estimate
$\delta \omega_2 \approx 2 \delta\Omega$ fits our data remarkably 
well.

Let us now return to the remaining moments $P_m^\rho$ shown 
in \Fref{fig:hmns_moments_amp}.
For all HMNSs, we observe the growth of $m=1$ and $m=3$
components. For the \texttt{SHT-M2.0-I} model, the $m=1$ perturbation 
becomes as large as the main $m=2$ mode at the end of the 
simulation. For the \texttt{LS220-M1.5} models, the HMNS 
starts collapsing before this happens. During the collapse, 
the $m=1$ moment is rapidly amplified.
We did not identify the exact nature of those components,
but similar results have been reported in \cite{Bernuzzi2013prd} for
BNS mergers of spinning stars employing an ideal gas EOSs.
Lastly, the $m=4$ pattern possesses the same angular velocity
as the $m=2$ mode, and both moments are, on average, decaying.
We therefore attribute the $m=4$ moment to the non-harmonic 
$\phi$-dependency of the $m=2$ oscillation mode, which in 
turn is due to the large amplitude.

A rather  surprising result is that the angular velocity 
$\dot{\phi} = \frac{1}{2} \omega_2$ of the $m=2$ mode pattern
is almost equal to the maximum angular velocity of the fluid 
for all equal-mass models, as one can see in \Fref{fig:hmns_rot_prof}.
There are two likely explanations besides pure coincidence.
Either the mode frequency is somehow determined 
by the maximum angular velocity for our models,
or the oscillation of the bulk affects the 
rotation rate of the outer layers. Animations of our 
simulations reveal strong nonlinear effects close to the 
surface, such as mass shedding. On the other hand, this
does not explain why the rotation rate decreases again
toward the surface.
In any case, this relation deserves further scrutiny in 
future studies.

\subsection{Gravitational waves}
\label{sec:gw}

In the following we will discuss the features of the gravitational
wave signal extracted from our simulations. The focus will 
be on relating the GW spectra to the dynamics of the HMNSs. 
For simplicity, we did not extrapolate the signals to infinity,
but use a finite extraction radius of $756 \usk\kilo\meter$
and $916 \usk\kilo\meter$ for the LS220 and SHT models, 
respectively.

We begin with the equal mass models.
In all cases, the GW strain multipole moments are dominated 
by the $l=m=2$ term, followed by the $l=2, m=0$ contribution, 
which has a maximum strain around 6--8\% of the former. 
The $m=3$ perturbation which develops for the HMNSs (see 
\Sref{sec:hmns}) is also visible in the $l=m=3$ moment of 
the respective GW signals, but the amplitude just amounts to 
1--2\% of the dominant mode.
The GW strain of the $l=m=2$ contribution is shown in 
Figs.~\ref{fig:gw_strain_sht} and \ref{fig:gw_strain_ls220}.
When comparing the spinning models to the nonspinning ones,
differences in the GW signal are visible by eye. Besides the 
orbital hangup during the inspiral, the modulation during the 
HMNS phase is different. We will come back to this issue later.

First, we try to understand the GW spectrum. 
As discussed in \Sref{sec:hmns}, the oscillation frequency of the 
main $l=m=2$ oscillation mode is strongly modulated due to the 
radial oscillation. This will at least broaden the corresponding
peak in the Fourier spectrum.
Considering that the signal spends more time in a given 
frequency bin near a local extrema of the modulated frequency, 
one might expect the appearance of additional side peaks 
located in proximity to the extrema. To explore the effects of 
frequency modulation on the Fourier spectra, we synthesize a toy 
model signal given by
\begin{align}
z &=
e^{ 
i \phi(t)
- \frac{t}{\tau_s}
} \\
\phi(t) &=
2 \pi f_s t - \frac{\Delta f}{f_m} \cos\left(2 \pi f_m t\right) 
   e^{-\frac{t}{\tau_m}}
\end{align}
This corresponds to a signal with a sinusoidal frequency modulation, where
both the signal and the frequency modulation amplitude decay exponentially,   
on different timescales. Figure~\ref{fig:synthetic_spec} shows the instantaneous 
frequency $\dot{\phi}$ and the Fourier spectrum of $z$, where we chose 
parameters approximating real signals. Indeed, we observe several peaks in the 
spectrum which are located near local extrema of the instantaneous frequency. 
We stress that this is not an exact relation, and not all local extrema 
cause separate peaks. Further, we even find a peak at a frequency larger 
than the maximum instantaneous frequency, although it is rather small.
Note the appearance of additional peaks due to frequency modulation is an 
entirely different effect than the so called combination 
frequencies, i.e., side peaks arising in Fourier analysis 
of monochromatic signals superposed in a non-linear fashion. 
We expect both effects in our results.

The effective strain
spectrum for the two irrotational models forming a HMNS is given
in the upper panels of Figs.~\ref{fig:specs_gw_hmns_sht} 
and~\ref{fig:specs_gw_hmns_ls}. 
In order to distinguish peaks that can be attributed to oscillations
of the HMNS from peaks caused by the plunge and merger stage,
we also show the spectrum of the late signal, starting 
$2\usk\milli\second$ 
after the time of the first density maximum during the merger.
The comparison shows that the dominant peak $f_2$ clearly originates 
from the HMNS phase in all cases.
The low frequency peak $f_1$ on the other hand is strongly 
suppressed in the late signal, and therefore has to be caused
by the plunge and merger. More precisely, it seems to originate
from the times of the plunge and the maximal expansion after 
the first 1--2 bounces. At this time the system still resembles 
more two separate cores in an eccentric orbit inside a common 
envelope than a deformed single star (compare 
\Fref{fig:polar_coords}).
We note that in \cite{Stergioulas2011b}, the low frequency 
peak (for different models) has been attributed to a combination 
frequency of the quasi-radial and $m=2$ mode oscillation of the HMNS. 
For the cases at hand, this interpretation does not apply. To 
distinguish between the two causes of the low frequency peak, one 
has to study the late GW signal separately.

\begin{figure*}
  \centering
  \includegraphics[width=0.65\columnwidth]{{{gw_strain_SHT-M2.0-I}}}
  \includegraphics[width=0.65\columnwidth]{{{gw_strain_SHT-M2.0-S}}}
  \includegraphics[width=0.65\columnwidth]{{{gw_strain_SHT-M2.2-I}}}  
  \caption{Gravitational wave strain at distance 100 $\mega\parsec$ 
  for models \texttt{SHT-M2.0-I} (left panel), \texttt{SHT-M2.0-S} (middle panel), 
  and \texttt{SHT-M2.2-I} (right panel). The waveforms were extracted at 
  radius $r = 916\usk\kilo\meter$.  
  }
  \label{fig:gw_strain_sht}
\end{figure*}

\begin{figure*}
  \centering
  \includegraphics[width=0.65\columnwidth]{{{gw_strain_LS220-M1.5-I}}}
  \includegraphics[width=0.65\columnwidth]{{{gw_strain_LS220-M1.5-S}}}
  \includegraphics[width=0.65\columnwidth]{{{gw_strain_LS220-M1.8-I}}}  
  \caption{Gravitational wave strain at distance 100 $\mega\parsec$
  for models \texttt{LS220-M1.5-I} (left panel), 
  \texttt{LS220-M1.5-S} (middle panel), and \texttt{LS220-M1.8-I} (right panel).
  The waveforms were extracted at radius $r=756 \usk\kilo\meter$.  }
  \label{fig:gw_strain_ls220}
\end{figure*}

To interpret the remaining peaks, we need to take into account
the modulation of the $l=m=2$ mode frequency.
For this, we juxtapose the instantaneous frequency of the $l=m=2$ 
component of $\Psi_4$
to the Fourier spectra in Figs.~\ref{fig:specs_gw_hmns_sht} 
and~\ref{fig:specs_gw_hmns_ls}.
The largest peak $f_2$ clearly corresponds to the average 
instantaneous frequency. 
Moreover, sidepeak $f_4$ 
in \Fref{fig:specs_gw_hmns_ls} is located at the same frequency 
as two of the minima of the instantaneous frequency. 
It is thus plausible that this peak is caused by the very same 
oscillation mode as peak $f_2$.

The high-frequency peak $f_3$ is located shortly below the global 
maximum of the instantaneous frequency, similar to what we found
for the synthetic signal discussed earlier.
The maximum frequency is reached during the 
merger when the star is most compact. For model \texttt{LS220-M1.5-I},
the maximum frequency is reached again during the second bounce.
For both models however, the distance of peak $f_3$ to the 
main peak \emph{also} agrees with the mean radial 
oscillation frequency (see \Tref{tab:hmns_freqs}).
For model \texttt{SHT-M2.0-I},
peak $f_3$ is barely visible in the spectrum of the late 
signal and should therefore be attributed to the merger phase
instead to a combination frequency.
For model \texttt{LS220-M1.5-I} on the other hand, peak $f_3$ is still
present in the late signal, although smaller. Therefore
it is most likely an overlap of the $l=m=2$ oscillation 
itself at the time of maximum compactness,
and a combination frequency of the same mode at the HMNS stage 
with the radial oscillation frequency.
The agreement of those frequencies is most likely a coincidence.
The shift of the maximum of the $l=m=2$ mode frequency relative to 
the central value depends on the amplitude of the radial 
oscillation during the merger, while the shift of the combination 
frequency is given by the frequency of the radial mode.

\emph{Our findings indicate that one cannot securely associate
such high frequency peaks either to the merger stage, a 
combination frequency, or even a separate HMNS oscillation 
mode. For this, one needs to consider the late part of the 
spectrum separately.
}
Of course, this requires a high signal-to-noise ratio.
We stress that the merger phase does not necessarily leave a strong
imprint in the GW signal. Figure~\ref{fig:specs_gw_hmns_uneq} shows
the GW spectrum and instantaneous frequency evolution for the 
unequal mass model \texttt{LS220-M1.4-U}. As one can see, there is no 
significant high frequency peak at all.
Also, the only low frequency peak is very broad and of low 
amplitude.
It seems to originate simply from the final part of the inspiral.

\begin{figure}
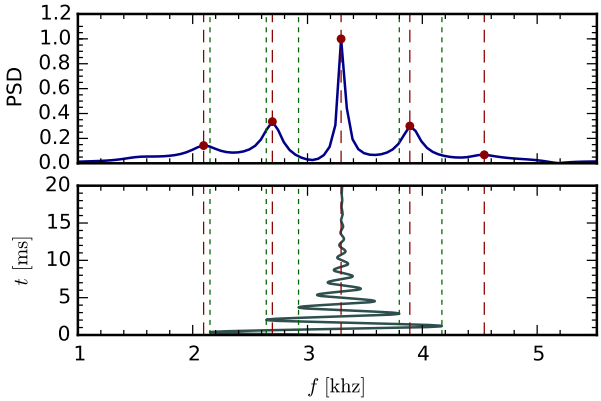

  \centering
  \includegraphics[width=0.95\columnwidth]{{{synthetic_spec}}}
  \caption{(Top panel) Fourier spectrum of toy signal, with values
  $f_s=3.3 \usk\kilo\hertz$, 
  $\Delta f= 1.3 \usk\kilo\hertz$,
  $\tau_s= 8 \usk\milli\second$,
  $ \tau_m = 3 \usk\milli\second$,
  $f_m=0.6\usk\kilo\hertz$.
  (Bottom panel) Corresponding instantaneous frequency.
  The vertical dashed lines mark the maxima of the Fourier spectrum,
  the vertical dotted lines the first five extrema of the instantaneous
  frequency.
  }
  \label{fig:synthetic_spec}
\end{figure}

\begin{figure}
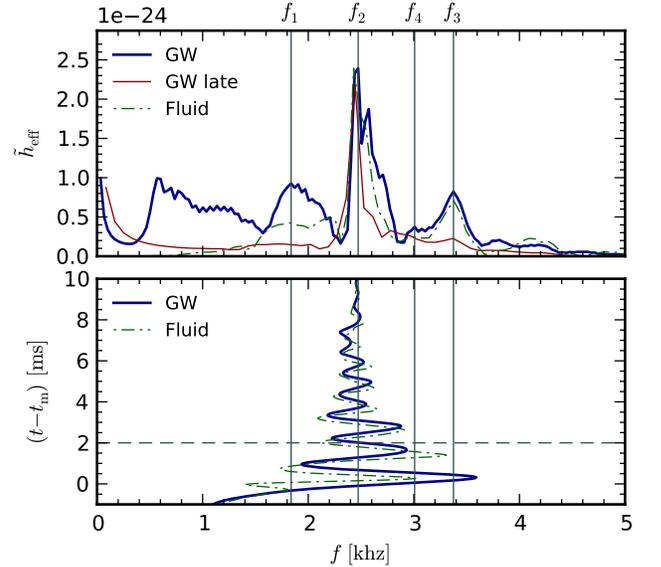

  \centering
  \includegraphics[width=0.99\columnwidth]{{{specs_gw_hmns_SHT-M2.0-I}}}
  \caption{(Top panel) GW effective strain spectrum for model \texttt{SHT-M2.0-I}.
  The thick blue curve shows the spectrum of the full signal at
  at nominal distance of $100\usk\mega\parsec$. The red solid line shows
  the spectrum of the late signal, starting $2\usk\milli\second$ (plus 
  time of flight) after the time of the first density
  maximum during the merger, $t_\mathrm{m}$.
  The green dash-dotted line depicts $f^2 \tilde{P}_2^\rho(f)$, where
  $\tilde{P}_2^\rho$ is the Fourier spectrum of the fluid moment 
  $P_2^\rho$, rescaled to match the maximum amplitude.
  (Bottom panel) Evolution of the instantaneous frequencies
  of the GW signal $\Psi_4$ (thick blue line), and the moment 
  $P_2^\rho$ (green dash-dotted line). 
  The GW signal has been shifted to compensate for the time 
  of flight.
  }
  \label{fig:specs_gw_hmns_sht}
\end{figure}

\begin{figure}
  \centering
  \includegraphics[width=0.99\columnwidth]{{{specs_gw_hmns_LS220-M1.5-I}}}
  \caption{Like \Fref{fig:specs_gw_hmns_sht}, but showing model \texttt{LS220-M1.5-I}.}
  \label{fig:specs_gw_hmns_ls}
\end{figure}

\begin{figure}
  \centering
  \includegraphics[width=0.99\columnwidth]{{{specs_gw_hmns_LS220-M1.4-U}}}
  \caption{Like \Fref{fig:specs_gw_hmns_sht}, but for unequal-mass 
  model \texttt{LS220-M1.4-U}.}
  \label{fig:specs_gw_hmns_uneq}
\end{figure}

In order to link the GW signal to the dynamics of the HMNS, we
also included the spectrum and instantaneous frequency of the fluid
moment $P_2^\rho$ in Figures~\ref{fig:specs_gw_hmns_sht} and 
\ref{fig:specs_gw_hmns_ls}.
The frequencies of the three main peaks agree well with the 
GW signal. Note the moments use a different radial weight than one
would use in the quadrupole formula to estimate the GW strength.
This might explain the different amplitude of the low 
frequency peak. Also, since the fluid mode phase velocity
depends on the radial weight unless there is only one dominant 
oscillation mode, the differences to the GW phase velocity
during the merger stage are to be expected.
Finally, the inspiral part is missing in the spectrum of 
$P_2^\rho$ because we only start computing it shortly before the merger.

\begin{figure}
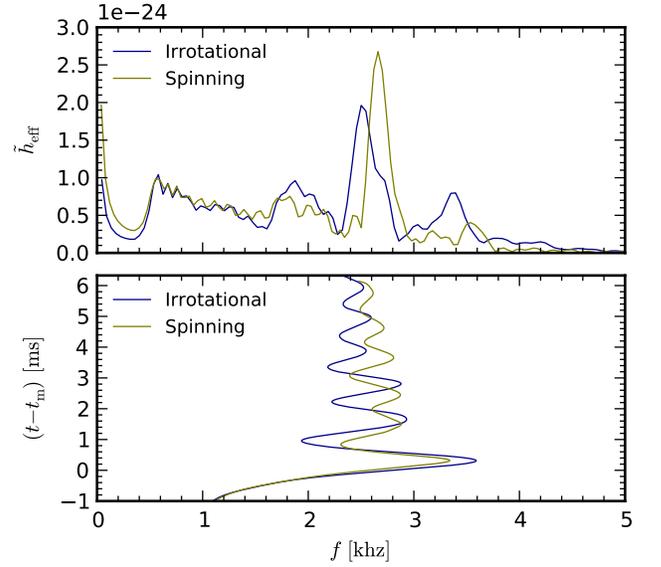

  \centering
  \includegraphics[width=0.99\columnwidth]{{{specs_gw_spin_SHT-M2.0}}}
  \caption{(Top panel) GW effective strain spectrum at nominal
  distance of $100\usk\mega\parsec$ for irrotational model \texttt{SHT-M2.0-I} 
  compared to the same spectrum for spinning model \texttt{SHT-M2.0-S}.
  The GW signal from the longer simulation was cut such that the 
  signal duration after the merger matches the shorter simulation.
  (Bottom panel) Evolution of the corresponding instantaneous 
  frequencies.
  }
  \label{fig:specs_gw_spin_sht}
\end{figure}

We now return to the influence of the spin on the waveform.
The spectra and instantaneous frequencies of spinning and 
irrotational models are compared in 
Figures~\ref{fig:specs_gw_spin_sht} and \ref{fig:specs_gw_spin_ls}
for the models with SHT and LS220 EOS, respectively.
Note that for cases where the HMNS does not collapse, 
the exact shape of the GW spectrum depends on the evolution time. 
A longer simulation enhances the contribution from the monochromatic 
late signal. In order to avoid such selection effects when comparing 
spinning and non-spinning GW spectra, we cut the GW signal
from the longer simulation such that the duration after the 
merger matches the shorter simulation. The relative peak amplitudes 
in Figs.~\ref{fig:specs_gw_hmns_sht} 
and~\ref{fig:specs_gw_spin_sht} are indeed different.

The frequency shift of the main peak for the spinning models
with respect to the irrotational ones is smaller than the 
width of the peak.
The main difference between the two cases is that the 
maximum instantaneous frequency
reached during the merger is slightly smaller for the spinning
models, which in turn is related to the weaker radial 
oscillation (see \Sref{sec:hmns}). Peaks which represent combination 
frequencies with the radial mode should thus have smaller amplitudes, 
while peaks representing the maximum frequency during the merger 
should be located at slightly lower frequency. Indeed, peak $f_3$ 
is split into two smaller peaks for the spinning LS and SHT EOS 
models, which further substantiates our interpretation.
Unfortunately, \emph{the dependency of the GW spectrum on the spin
is a complicated one, and it would be extremely difficult 
to infer the spin from spectra} like the ones presented here.

\begin{figure}
  \centering
  \includegraphics[width=0.99\columnwidth]{{{specs_gw_spin_LS220-M1.5}}}
  \caption{Like \Fref{fig:specs_gw_spin_sht}, but for models 
  \texttt{LS220-M1.5-I} and~\texttt{-S}.}
  \label{fig:specs_gw_spin_ls}
\end{figure}

\subsection{Matter ejection}
\label{sec:ejecta}
%

Finally, we study the mechanisms leading to the ejection 
of matter during and after the merger,
in particular the amount of unbound matter.
The models undergoing prompt collapse have no significant disk
and no unbound matter. Hence we only discuss the ones
producing a HMNS.

\begin{figure}
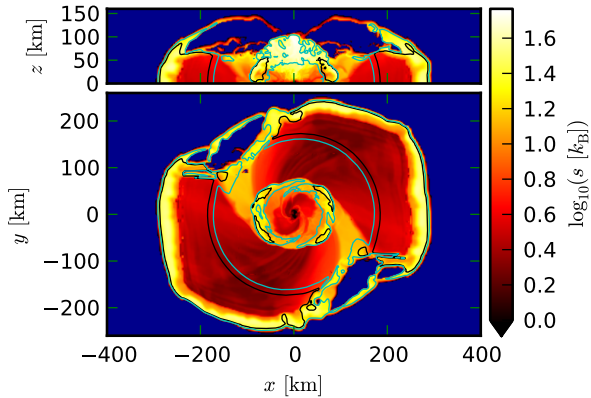

  \centering
  \includegraphics[width=0.99\columnwidth]{{{entropy_xyxz_SHT-M2.0-I}}}
  \caption{Specific entropy in the orbital plane, $2.5 \msec$ 
    after the merger, for model \texttt{SHT-M2.0-I}. The regions where 
    matter is unbound according to the geodesic criterion are
    bounded by the black lines, the cyan lines enclose regions
    unbound according to the Bernoulli criterion. Regions covered 
    by the artificial atmosphere are colored blue.}
  \label{fig:entropy_xyxz}
\end{figure}

\begin{figure}
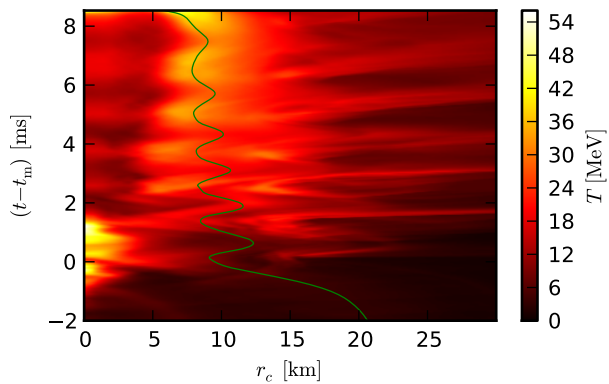

  \centering
  \includegraphics[width=0.99\columnwidth]{{{rcirc_t_m0_temperature_LS220-M1.5-I}}}\\
  \caption{Thermal evolution of model \texttt{LS220-M1.5-I} from merger 
  until BH formation.
  The color corresponds to the $\phi$-averaged temperature in the 
  orbital plane as a function of circumferential radius and 
  coordinate time. The green line marks the density averaged 
  circumferential radius $\bar{R}_c$.}
  \label{fig:rcirc_t_temp}
\end{figure}

A significant amount of matter is ejected from the remnant 
during the merger, but also at later stages. 
A large part of the matter ejected during the merger forms
an envelope/disk which does not become unbound. 
The main effect leading to unbound matter for our equal-mass 
models seems to be shock formation.
The $m=2$ mode oscillation 
is highly nonlinear at low densities, causing two spiral shaped shock 
waves, which in turn liberate matter from the disk.
Effects like this have also been observed by~\cite{Hotokezaka2013}
for various equal-mass models with hybrid EOSs.
The shape of the expanding shock fronts in the orbital and $xz$-planes 
is illustrated in \Fref{fig:entropy_xyxz}.
Moreover, the strong radial oscillation modulates the strength of
shock formation,
such that each wave of ejected matter can be traced back to a
radial oscillation (or the merger itself).
The evolution close to the NS is pictured in \Fref{fig:rcirc_t_temp},
displaying a spacetime diagram of the $\phi$-averaged temperature
in the orbital plane. Outside a radius $\approx 15\usk\kilo\meter$,
one can easily spot outgoing waves of increased temperature.
The comparison with the average radius (marked by the green line)
reveals that they are clearly correlated with the radial oscillation.

In order to compute the amount of matter that will escape to infinity,
we integrate the density of unbound matter over the whole computational 
domain. There are two options for estimating whether or not a fluid 
element is bound. Both assume a stationary spacetime and coordinates 
such that $t^\mu$ is a Killing vector.
The first criterion, which is also used in~\cite{Hotokezaka2013}, 
is based on the assumption that fluid elements 
move along geodesics. This implies that the 4-velocity $u^\mu$ 
satisfies $u_t = \text{const}$. At infinity, $u_t = -W$, where $W$ 
is the Lorentz factor. A fluid element would hence be able to reach 
infinity if $-u_t > 1$, approaching an escape velocity of
$v_e^2 = 1-u_t^{-2}$. This criterion is only meaningful if the 
pressure forces stay small compared to the gravitational ones. In 
the context of NS mergers, this is not the case near the remnant. 
Pressure forces mainly accelerate material outwards, such that it 
may become unbound at a later time. The geodesic criterion 
tends to underestimate the amount of unbound matter. We use it to 
obtain a lower limit.

The second option is to assume a stationary relativistic fluid 
flow, for which $h u_t$ is conserved along fluid worldlines.
We note that the relativistic enthalpy $h$ is only defined 
up to a constant factor that depends on the formal baryon mass 
used to define the rest mass density $\rho$ in terms of the baryon 
number density. In our simulations, we chose this constant such 
that $h \to 1$ when density and temperature tend to zero. We then 
assume that this limit is approached by the fluid at infinity, and 
arrive at the Bernoulli criterion $-h u_t > 1$. 
The Bernoulli criterion might overestimate the role of internal
energy for liberating matter. In our simulations, shock formation 
is an important cause for matter ejection. The corresponding shock 
heating can increase the value of $h$, such that the Bernoulli 
criterion is satisfied. However, the assumption of a stationary 
fluid flow is a decidedly bad approximation near those shocks. 
In any case, the Bernoulli criterion generally predicts more 
unbound mass than the geodesic criterion. 
We note there is a third estimate used in 
\cite{Oechslin02, Bauswein2013b}. Contrary to \cite{Oechslin02},
it is not the geodesic criterion, but in general closer to
the Bernoulli criterion. It can be reformulated as
$-h u_t > 1 + \frac{\alpha P}{W \rho}$, where $\alpha$ is the
lapse function.

A comparison between the Bernoulli and geodesic criteria is shown 
in \Fref{fig:entropy_xyxz}. They agree well for the fluid in first
wave of ejected material, which at this point is already very diluted.
Closer to the remnant however, where the density is also higher, 
one can spot large differences
near the spiral shocks. Based on the data in the orbital plane
at the time shown in \Fref{fig:entropy_xyxz},
we estimate that the Bernoulli criterion yields at least two times
more unbound matter than the geodesic one for model \texttt{SHT-M2.0-I}.
For the unequal-mass model \texttt{LS220-M1.4-U}, we even find
a ratio larger than five. In principle, the ratio could become 
arbitrary high in cases where the material is just marginally unbound 
according to the geodesic criterion. Also note the comparison in the 
orbital plane might not be representative for the full volume.

Another source of uncertainty is that the fluid is set to
the artificial atmosphere wherever the density drops below
$6 \times 10^7 \usk\gram\per\centi\meter\cubed$ (close
to the lowest tabulated density of the SHT EOS).
At the time shown in \Fref{fig:entropy_xyxz}, the first wave of 
ejected matter already exhibits patches covered by artificial 
atmosphere. Clearly, it would have been beneficial to use a 
lower density for the artificial atmosphere, extending the 
EOS if necessary.
Another possible improvement for future simulations would
be to use the time integral of the unbound matter \emph{flux} 
through a sphere with a radius sufficiently large to avoid
using criteria for boundness in highly dynamic regions.

\begin{figure}
  \centering
  \includegraphics[width=0.99\columnwidth]{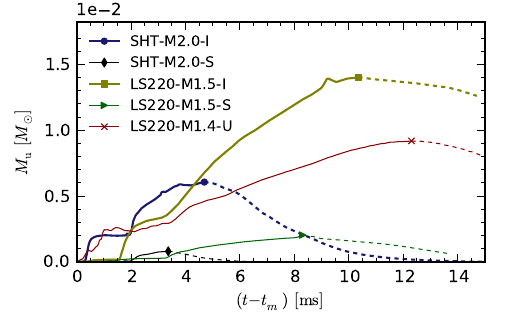}
  \caption{Time evolution of the estimate for the mass of unbound matter, 
  according to the geodesic criterion. The decreasing parts (dotted 
  lines) are unphysical artifacts caused by the artificial atmosphere,
  the maximum serves as a lower bound for the real value.}
  \label{fig:ejecta_unbound_mass.}
\end{figure}

Despite the above uncertainties, we can use the geodesic criterion 
to get a lower limit on the unbound mass. The time evolution of the 
formally unbound mass is shown in \Fref{fig:ejecta_unbound_mass.}.
Note that although the steepest increase occurs shortly after the 
merger, matter becomes unbound continuously, driven by the shock 
waves caused by the NS oscillations.
The decrease at late times is a numerical artifact caused by matter 
diluted to artificial atmosphere density. We use the maximum as 
a lower limit for the total amount of matter that would escape 
to infinity. In addition, we evaluate the density-weighted average 
specific entropy and electron fraction, as well as the maximum escape 
velocity at the time this maximum occurs. The results are reported 
in \Tref{tab:ejecta}. Note however that the electron fraction
is passively advected and would in reality be changed by neutrino
radiation of the hot and optically thin fluid.
Model \texttt{SHT-M2.0-I} is also investigated in \cite{Bauswein2013b},
using the conformal flatness approximation.
They report an ejected mass of $9.08 \times 10^{-3} \usk\Msol$,
which is compatible with our lower limit.
Surprisingly, the amount of unbound matter for the spinning 
models is markedly smaller compared to the irrotational ones.
We attribute this to the different amplitude of the radial 
oscillations shown in \Fref{fig:hmns_radius_evol}. 
\emph{At least for equal-mass models, the initial NS spin 
can have a strong influence on the amount of matter ejected 
to infinity.}

\begin{table}
\begin{ruledtabular}
\begin{tabular}{lcccc}
Model&
$M_u\, [0.01\usk\Msol]$&
$\bar{s} \, [k_\mathrm{B}]$&
$\bar{Y}_e$&
$v_e$
\\\hline 
SHT-M2.0-I&
$0.61$&
$18$&
$0.11$&
$0.33$
\\
SHT-M2.0-S&
$0.08$&
$23$&
$0.10$&
$0.44$
\\
LS220-M1.5-I&
$1.40$&
21&
$0.14$&
$0.28$
\\
LS220-M1.5-S&
$0.20$&
$18$&
$0.09$&
$0.24$
\\
LS220-M1.4-U&
$0.92$&
$16$&
$0.10$&
$0.19$
\end{tabular}
\end{ruledtabular}
\caption{Properties of unbound matter. $M_\mathrm{u}$ is the lower 
limit for the unbound mass, obtained from the maximum amount of unbound 
matter according to the geodesic criterion. At the time the maximum 
occurs, we compute the density weighted average of specific entropy and 
electron fraction, denoted by $\bar{s}$ and $\bar{Y}_e$. $v_e$ is the 
maximum velocity unbound material would reach at infinity.}
\label{tab:ejecta}
\end{table}

\section{Summary and Discussion}
\label{sec:summary}

In this work we have investigated in detail the merger and post-merger
dynamics of irrotational equal-mass BNS models in comparison to models 
with initial NS spins aligned to the orbital angular momentum.
The rotation rates considered are more than four times lower 
than the fastest observed pulsar, but still larger 
than any observed pulsar in a BNS system.

We have found that the spin affects the amplitude of the post-merger radial 
oscillations. The reason might be a slower impact due to the orbital 
hangup effect described in \cite{Tsatsin2013,Bernuzzi2013prd}, which we 
observe as well during the inspiral. The additional angular momentum 
might also contribute by increasing the centrifugal barrier. The 
radial oscillation is noteworthy because it turns out to affect the 
GW spectrum as well as the amount of unbound matter.

The GW spectrum is affected by the radial oscillation because 
it leads to a modulation of the rotation rate. Furthermore,
the main $m=2$ oscillation mode frequency in the inertial frame
is correlated with the rotation rate, and hence shows
variations as well. The modulation of the frequency can become 
strong enough to create additional peaks in the GW spectrum.
In particular, the frequency maximum reached when the star is 
most compact during the merger leads to a separate high-frequency 
peak for the equal-mass systems we studied.
We also observed that local extrema of the instantaneous mode 
frequency can cause additional side peaks of the dominant peak.
The frequency modulation and its influence on the GW spectrum 
was already described in \cite{Hotokezaka2013c}. In addition, 
we found the quantitative relation 
$2 \Delta F_\mathrm{R} \approx \Delta F_2$
between the modulation amplitudes of the average rotation rate 
and the $m=2$ mode frequency. It is yet unknown if this relation 
holds in general.

The direct influence of the spin on the frequency of the $m=2$
oscillation is weak for our models, comparable to the width of 
the corresponding peak. Inferring the spin from the GW signal
will prove difficult. On the other hand, a weak dependency on spin
is advantageous when using the GW spectral features to break the 
mass-redshift degeneracy of the inspiral GW signal, as proposed
in \cite{Messenger2013}.

The appearance of additional peaks due to frequency 
modulation should not be confused with the so called 
combination frequencies, which can be present simultaneously.
Indeed, we observed that the peak due to the maximum frequency 
of the main $m=2$ mode reached during merger can overlap with 
a combination frequency between the post-merger $m=2$ and radial 
modes. The low-frequency peaks on the other hand are dominated 
by the contribution from the plunge and the first one or two 
bounces for our models. At those phases, the system
still exhibits a double core structure, which suggests that it
might be better described in terms of orbiting cores 
with very strong tidal effects instead of interpreting it as 
one strongly oscillating star. This view is also supported 
by the discovery of EOS independent empirical relations for 
the low frequency GW signal from the merger phase, see 
\cite{Bernuzzi2014prl, Takami2014}.

We stress that the above findings obtained for our selected models,
which are all very massive, are not to be generalized. For 
example, the unequal-mass model we evolved shows only small 
modulation and no significant high- or low-frequency peaks.
A large number of BNS mergers has been investigated in 
\cite{Hotokezaka2013c}.
Some of those models led to a large modulation similar to the
ones found in our simulations, while others showed a 
weak modulation (or drift), resulting only in the broadening 
of the main peak. 
Nevertheless, a consequence for GW analysis is that one 
needs to study the spectrum of the late signal separately in 
order to determine the nature of the high-frequency peaks.
The interpretation as combination frequencies proposed in
\cite{Stergioulas2011b} does not hold in general.

Another important aspect of the merger, besides the GW signal,
is the amount, composition, and temperature of the matter ejected
from the system. For our models, matter is liberated in several
waves along spiral shock waves originating from the HMNS. The 
first wave is launched during the merger itself. The subsequent 
high-amplitude $m=2$ oscillation causes a shock wave in the 
form of a double spiral. The initial strength is modulated by 
the radial oscillation. We observe that more material becomes 
unbound for the irrotational models, which is most likely due 
to the stronger radial oscillations we observe  for this case.

One of our irrotational models was also evolved by 
\cite{Bauswein2013b}. Their estimate for the amount of ejected 
matter, which is based on a prescription similar to the Bernoulli 
criterion, is compatible with our lower limit based on the geodesic
criterion. However, we found that the Bernoulli criterion can 
predict an unbound mass several times higher than the more 
conservative geodesic criterion. The differences are mainly 
caused by the usage of boundness criteria which assume some form 
of stationarity in regions which are still highly dynamic. 

All our models are close to the threshold mass for prompt 
collapse to a BH, which was computed for a range of EOSs 
in \cite{Bauswein2013} using the conformal flatness approximation, 
and which we could confirm in full GR for the LS220 and SHT EOSs 
considered here. Given that those masses exceed the maximum mass 
for a uniformly rotating star, we need to ask what prevents the 
collapse. As it turns out, the standard notion that the prompt
collapse is prevented by a differential rotation profile where the core 
rotates more rapidly does not apply to our models. Instead, we 
find an extended envelope rotating slightly below Keplerian velocity, 
while the core rotates more slowly than the outer layers. 

We stress that envelopes are not necessarily responsible for
delaying the collapse in general. However, their potential 
presence is an important aspect to be considered when studying 
the long-term stability of merger remnants.
In cases where the system is stabilized not mainly by 
rapid rotation or temperature of the dense core, it might be 
less sensitive to angular momentum loss or cooling of the core.
Conversely, neutrino cooling and magnetic fields might 
have a stronger impact on the envelopes than on the dense 
cores. Of course, the impact of thermal effects on the core 
depends on the EOS and the mass. Using a simplified analytic 
EOS and an idealized cooling mechanism, the BNS merger 
simulation presented in \cite{Paschalidis2012} suggests 
that cooling can reduce the lifetime of HMNSs.
For nuclear physics EOS, the influence of thermal effects 
on the stability of isolated massive stars tends to be weak, 
see \cite{Galeazzi2013, Kaplan2013apj}. On the other hand, those 
studies do not take into account extended envelopes, leaving 
room for thermally induced collapse for such systems.

\begin{acknowledgments}
We would like to thank A.~Bauswein, H.T.~Janka, N. Stergioulas, and K. Takami
for helpful discussions, and especially thank J. A. Font for 
carefully reading the manuscript and help with the introduction.
Further, we acknowledge discussions with L. Rezzolla on the project,
in particular on the HMNS density profile.
This work was granted access to the HPC resources of  the Rechenzentrum 
Garching (\texttt{Hydra} supercomputer)
made available within the Distributed
European Computing Initiative by the PRACE-2IP, receiving funding from the
European Community's Seventh Framework Programme (FP7/2007-2013) under
grant agreement number RI-283493.
Part of the numerical computations were also carried out on the
  \texttt{Datura} cluster 
at the AEI. W.K. was supported by the Deutsche Forschungsgemeinschaft 
Grant SFB/Transregio 7.
F.G. was supported by the Helmholtz International Center for FAIR 
within the framework of the LOEWE program launched by the State of Hesse.
\end{acknowledgments}

\bibliography{aeireferences}

\end{document}